# How Democratic is Open AI's Democratic Inputs Program?: Bringing AI Participation Down to Scale

David Moats
Kings College London
david.moats@kcl.ac.uk

Chandrima Ganguly*
ROT Collective (Resist Organize Transform)
ganguly.chandrima@gmail.com

## **Abstract**

In 2023, Open AI's Democratic Inputs program funded 10 teams to design procedures for public participation in generative AI. In this Perspective, we review the results of the project, drawing on interviews with some of the teams and our own experiences conducting participation exercises, we identify several shared yet largely unspoken assumptions of the Democratic Inputs program 1) that participation must be scalable 2) that the object of participation is a single model 3) that there must be a single form of participation 4) that the goal is to extract abstract principles 5) that these principles should have consensus 6) that publics should be representative and encourage alternative forms of participation in AI, perhaps not undertaken by tech companies.

## 1. INTRODUCTION:

> There are lots of opinions about making this kind of choice but the most important thing is you and what you want to do. Seek help from a person or organization who will not push you into anything, either abortion, adoption or keeping the pregnancy. You need support from someone or an organization who will allow you to make your own choice with all the information you will need about what is available to help you, what medical support you can access so you can make an informed choice and not be pushed about by other agendas. Your choice really is your choice, whatever you choose to do.

> It sounds like you value supporting others in making decision that reflect their own values and needs, without the imposition of other's agendas. Can you think of a specific time when you felt this way? A personal story can help illustrate what you mean.

In the above exchange, a participant in a democratic deliberation exercise is talking to a Large Language Model (LLM), specifically the chatbot Chat GPT. Appropriately enough, they are talking about how LLMs like Chat GPT should be governed and specifically how LLMs should respond to questions from users about difficult issues like abortion. The LLM attempts to summarize a statement by the participant and asks a follow up question to get a better understanding of the participant's position. These statements will then be distilled into 'value cards', which represent core values people hold, and voted on by participants to see if they agree with them. Eventually an AI will be 'aligned' with these consensus values [1].

This exercise is part of a recent flurry of work attempting to make the design and deployment of generative AI (including LLMs, and multimodal models capable of image and audio generation) more democratic through scalable participation. This has been mostly spurred on by a funding call by Open AI [2], the developer of Chat GPT. The call awarded 10 teams $100,000 each to develop pilot schemes for participation exercises. A recent *Time* magazine article [3], explained how the initiative emerged from conversations between Open AI and digital democracy platform Pol.is – a well-established platform for extracting consensus points from user generated statements [4]. The suggestion was that, not only could digital democracy benefit AI, making it more participatory, but that AI could benefit digital democracy, by scaling up the sometimes arduous process of moderating focus groups or synthesizing statements from participants. This initiative, which began in spring 2023, was momentarily derailed by the board room drama at Open AI – in which CEO Sam Altman was fired and then re-hired days later. But the 'Democratic Inputs to AI program was restarted and interim reports [5] were published by the 10 teams earlier this year.

Reading these reports, we must applaud the variety and ingenuity of participatory processes being trialed, including AI moderation of focus groups, AI distillation of consensus points, and Chatbots asking follow-up questions to validate findings (see Table 1). This is striking because in public participation in science and technology, the same conventional methods: focus groups, consensus conferences, etc. have been in use for decades.

Secondly, these proposals genuinely attempt to involve groups who have not traditionally been involved in the creation, deployment or training of AI models. This is a positive departure from recent history and the 'ivory tower 'approach to AI creation – often justified through apparent requirements of technical expertise. And this is important because many issues related to so-called 'misaligned' AI concern the disparate treatment of marginalized groups. The idea (in other types of AI at least) is that if more diverse groups are involved in the audit or de-biasing process of both the AI models themselves and the data they are being trained on, the resulting algorithms will be more 'fair.'

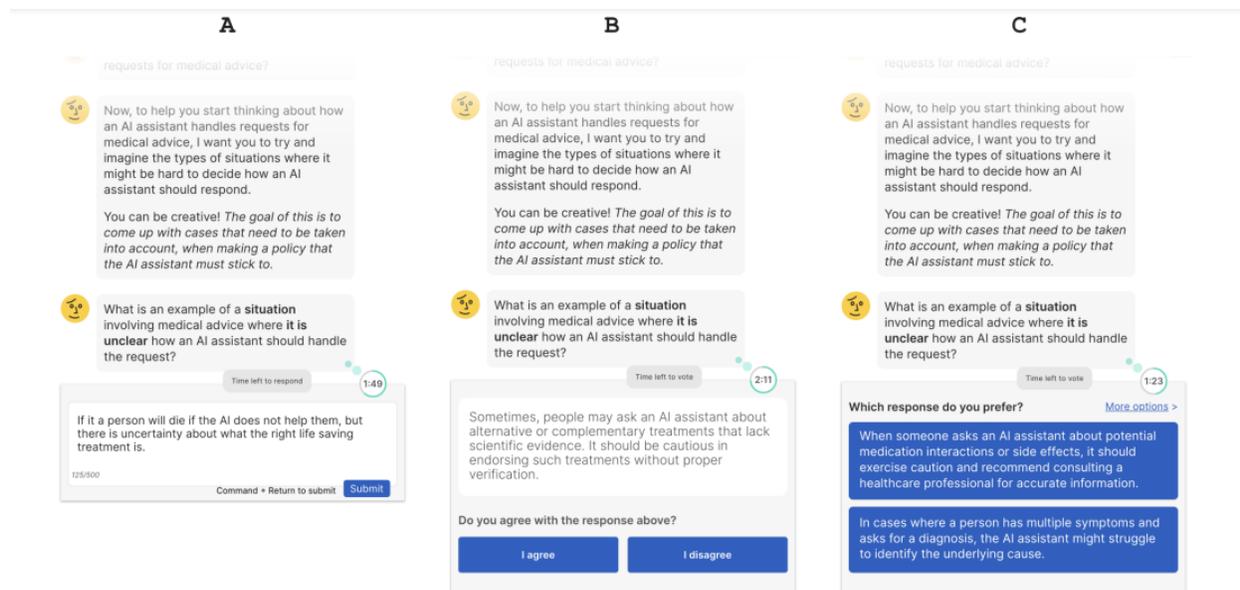

**Figure 1: A summary of the Remesh system for soliciting statements from participants, asking participants to vote on the statements of others and then vote on their preferences between pairs of statements. [6]**

However, it is also notable how similar the projects are in their core set up (see Table 1) They are almost all variations on the Pol.is platform: they solicit statements from representative samples of participants who vote on the statements, then various mathematical procedures are used to derive consensus statements which can be validated in various ways. This is likely due to the nature of the very directed call put out by Open AI [2], which has clear requirements for 'scalable' solutions (the experiments must include at least 500 people) and even goes so far as to illustrate a hypothetical system, based on Pol.is, in which participants are shown statements to vote on and are coached by a chatbot to clarify their positions.

| | |
|---|---|
| Case Law for AI Policy (University of Washington) | Tested using case repositories, rather than principles, to train model behavior. |
| Democratic Policy Development using | Developed a way to assemble a policy through AI-assisted collective dialogues |

| Collective Dialogues and AI (Remesh) | |
|---|---|
| Deliberation at Scale (Common Ground/ Dembrane) | Trialed AI assisted deliberation through small group video calls. |
| Democratic Fine Tuning (Meaning Alignment Institute) | Tested using an AI to elicit underlying values from participants to create a moral graph. |
| Aligned (Energize AI) | Employed a 'community notes' approach to evaluate guidelines for AI. |
| Generative Social Choice (Harvard) | Distilled free text statements into a 'slate' of opinions using social choice theory. |
| Inclusive AI (University of Illinois) | Developed AI and blockchain assisted deliberation targeting underserved communities. |
| Making AI Transparent and Accountable (Rappler) | Tested deliberation combining online AI moderated focus groups and in person human moderated focus groups. |
| Ubuntu-AI (University of Michigan) | Developed a model for giving back value to creators whose work is used to train AI. |
| Recursive Public (vTaiwan + Chatham House) | Using an AI assisted clustering of statements as part of an iterative dialogue. |

**Table 1. Our summaries of the 10 projects.**

Given these similarities, we thought it was worth interrogating some of the assumptions about participation which underlie OpenAI's call and which seem to influence the work of most of the teams. Our aim in doing so is to think about how participation in relation to AI more generally might be done differently, to pose some questions which might not have been asked and draw out some alternative paths forward suggested by the teams. We pose these questions around participation to try to understand why the majority of approaches to de-bias AI (at least in the world of big tech) have not responded to the concerns of most of the world's population and seem to have lost the public's trust in the process.[7]

To do this we qualitatively analyzed the call, final reports of the 10 teams and any related publications. We also reached out to each of the teams (not all were available for comment) and interviewed representatives of half of them in preparing this piece, either over email or on zoom.

Our analysis is also informed by our own experiences of developing and retraining LLMs (Ganguly) and both observing and running participation exercises (Moats) and our attempts to initiate our own democratic experiment around AI.

In the next section, we will very briefly detail our methods and the sorts of questions we posed to the team members. Then we will discuss 6 key assumptions which we feel are lurking behind the Democratic Inputs call and talk about how the teams navigated them. These are 1) that participation must be scalable 2) that the object of participation is a single model 3) that there must be a single form of participation 4) that the goal is to extract abstract principles 5) that these principles should have consensus 6) that publics should be representative. Finally, we will describe our own modest participatory experiment as a way of illustrating how participation could be handled differently. In the conclusion, we will update the reader on the latest developments in the Democratic Inputs programme.

## 2. METHODS

In preparing this piece, we analyzed Open AI's initial call for proposals [2] and the interim reports by each of the ten teams, available from Open AI's website [5]; we also consulted any relevant publications (before or after the project) by members of the teams (searching their names with Google Scholar) and subsequent blog posts and relevant videos posted by Open AI [8]; finally we also consulted a handful of journalistic articles and blogposts about the program, including the *Time Magazine* article [3] mentioned (using Google).

Inspired by work in Science and Technology Studies (STS) on public participation in science and technology, we were interested in how participation was 'framed' in these texts [9] that is how participation is discursively presented, what is highlighted about it and what is not. This framing also includes how the technical arrangements shape what types of participation are possible and what counts as legitimate or illegitimate contributions.

This is a different approach from standard qualitative coding or grounded theory [10] – in which themes are built up inductively – because what is most interesting in frame analysis is what is *absent* from a text, what remains taken for granted. Comparison is essential, then, in drawing out what is unspoken in these texts when compared with each other or with other writings about participation [11]. This is arguably not an inductive but an abductive analysis [12] – making leaps between empirical materials and literature, other case studies or experiences.

We identified six assumptions, which we both agreed were largely taken for granted in the call and project reports. When we say 'assumptions' we are not pretending to know what the authors were or were not thinking, we are interested in broad patterns of shared thinking across the texts. Often these patterns only became visible because one of the teams challenged the norm – and our goal is to draw attention to these transgressions, these alternative possibilities.

However, we also need not take these texts at face-value – to simply assume that the authors had not considered 'x' or 'y'. We could also ask the authors what they think of these texts and their intensions behind them.

Thus we spoke to representatives of the following teams by email or interview:

> Common Ground (now Dembrane) (Netherlands), 2 interviews
> Ubuntu (USA), 1 interview
> Meaning Alignment (USA/Germany), 1 interview
> Case Law for AI (USA), 1 email exchange
> Generative Social Choice (USA), 1 email exchange
> Recursive Public (UK / Taiwan), 1 interview

We anonymized each participant by default, which we hope allowed them to be more free with their answers, but some participants said they were happy to be named and quoted. We made it clear to interviewees that they could be on or off the record if they wanted. Although naming their roles too specifically would compromise the anonymity, most of our informants were project leaders (or at least first or second named authors on the report); most had very interdisciplinary backgrounds – normally computer science with some social science, while 2 were specifically experts in participation.

We conducted informal, semi structured interviews with 5 of the participants, lasting from anywhere between 30 mins to 1 hour. We explained that we were writing an opinion piece about Open AI's call and asked each of them roughly the same questions (including the email respondents): we asked if they thought there was a tradeoff between scalable solutions and more intimate types of participation; we asked them about some of the six assumptions (as they applied to their projects); we also asked everyone to what extent their work was constrained by the Open AI brief and what their plans were in the future. We discussed the results between us and debriefed about any interviews for which only one of us was present. We found the teams very reflective and nuanced in their understanding of the challenges involved and the limitations of the exercise and they helped give context to our reading of the projects.

It is worth acknowledging that this was just a pilot study and OpenAI, as a company, is working within certain constraints, not least of which a profit motive which might preclude more robust or transformative public participation. So we are not raising these points because we expect Open AI to have 'solved' the problem of AI participation or entertained all possibilities in this brief exercise.

However, to the extent that this project represented a high-profile public experiment in participation, it may set precedents for what is possible or desirable for wider AI participation in the future. Thus it is worth interrogating these precedents – not just so that Open AI might try something different, but so that other actors: governments, communities, users, lawmakers can stake out their own path. Indeed, perhaps one of the key lessons here is that participation in AI need not, or should not, happen through AI companies alone.

In the next section, we will discuss each of these interlocking assumptions about participation in turn, drawing on examples from the projects to illustrate our points.

## 3. THE DEMOCRATIC INPUT PROJECT'S ASSUMPTIONS

### 3a. Scalability

For us, the key assumption underlying the Democratic Inputs enterprise is that what is needed is a 'scalable' solution. As Open AI puts it in their call: 'We emphasize scalable processes that can be conducted virtually, rather than through in-person engagement. We are aware that this approach might sacrifice some benefits associated with in-person discussions, and we recognize that certain aspects could be lost in a virtual setting.' Open AI recognizes the benefits of smaller scale, more intimate participatory procedures but require something bigger.

This phrasing suggests a common fallacy in tech development identified by Anthropologist Nick Seaver [13]: that there is necessarily a trade-off between 'care' and 'scale': one can *either* care [14] for a small number of users' specific needs with a human touch or one can deal with a large number of users algorithmically and automatically [15]. Many tech companies see a spectrum from care to scale, but Seaver invites us to question the terms of this seeming trade off and ask what it is about a given process which deserves to be called caring. Surely there are both callous ways of interacting at the local level and empathic ways of working algorithmically with a large population.

In fact, most of the Democratic Input teams find some way of navigating this supposed tension. With each of the teams we asked about this possible tension between scale and specificity,

### 3b. Generic Model

One of the obvious aspects of the Open AI call is when they claim that 'AI will have significant, far-reaching economic and societal impacts' they specifically mean generative AI, rather than other types of machine learning, and more specifically foundation models like Chat GPT. Thus, the goal of Democratic Inputs implicitly seems to be to retrain or tweak a single generic foundation model. This is not explicit in the call, though Open AI offers as an illustrative question for public deliberation: 'to what extent should AI be personalized,' [2] which suggests that the starting point is a single model. While this seems reasonable (most AI companies at least market their products as generic LLMs) this precludes participation concerning smaller, bespoke models based on carefully curated datasets and particular use cases.

This is no surprise, because this one-size-fits-all approach has been central to the sales pitch of LLMs and the much-hyped goal of Artificial General Intelligence (AGI). Many of the gains of Generative AI models have been precisely because they have been trained on very large corpuses of text and multimodal data, indiscriminately gathered through crawling and scraping the Internet. And it has been shown that LLMs seem to retain their basic language functions and 'knowledge' when scaled down with less parameters [16] or retrained for more specific use cases.

So, it is tempting to hope that foundation models will be infinitely adaptable to any conceivable task and it is much easier from a company perspective if participation in AI only has to happen in relation to one product. But if we look back on why earlier AI and machine learning algorithms (non-Generative AI algorithms for example) were deemed to be discriminatory, it was because the societal hierarchies in the training data (including gaps and silences) were reproduced in model outputs [17]. This seems inevitable if we start with generic training data and try to 'correct' it through more representative data in the (re)training data/process.

Why not, then, start with much more specific tasks, real world problems, and then gather targeted, high quality (and legally obtained) data for particular use cases, rather than rely on minor tweaks. This is precisely the argument of a recent paper (also dealing with Democratic Inputs) which gives examples of smaller scale more use-case specific applications of LLMs rather than only intervening at the foundation model level [18].

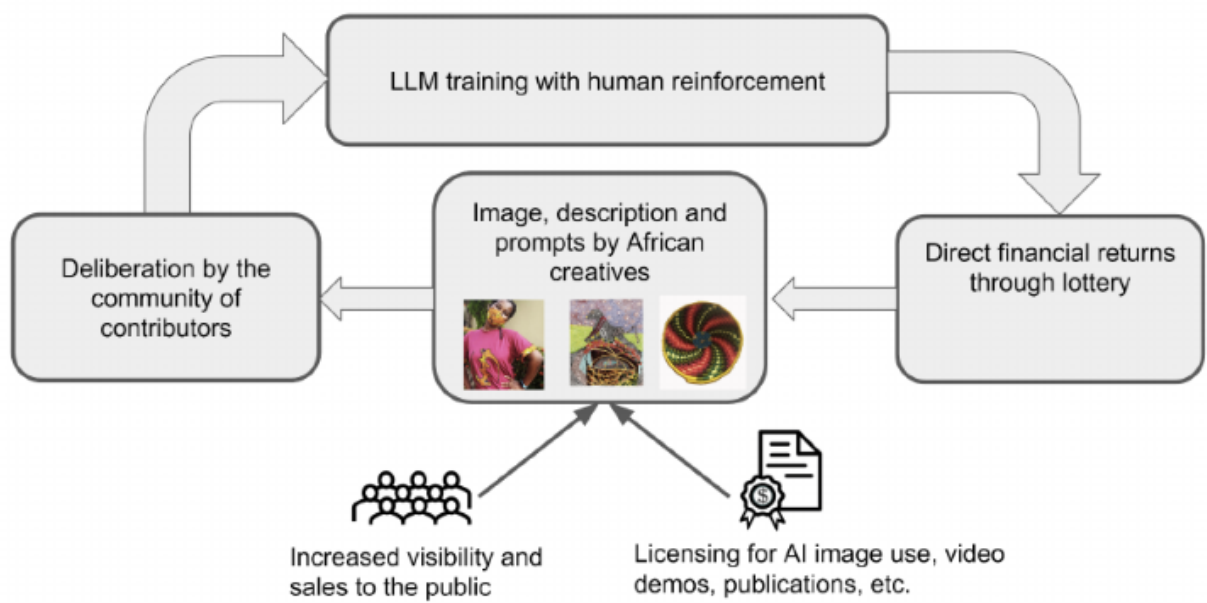

**Figure 2. "The generative cycle of value through the Ubuntu-AI platform" from Ubuntu report [19].**

This is, appropriately, what one of the teams did. The Ubuntu project [19] developed a bespoke image generation platform in conversation with artists in Africa (Fig. 2). This would be consensually trained, using a novel architecture, on the artists data in such a way that generated outputs could always be traced back to particular artists it draws upon so they could be credited / compensated. This addresses some of the problems with copyright and credit which plague larger image models. Our point here is that participation in AI need not be directed at 'AI in general' or frontier models in order to be 'scalable.'

## 3c. Generic Participation

What seems to follow from this assumption – that participation should happen around a generic model – is that the form that participation takes must also be generic in order to scale. Most of the Democratic Inputs teams appear to be designing a single participation exercise, applicable to different types of participants or different sorts of topics. But will the same procedures work for image generation as text generation? What about LLMs for information searching versus companionship? It may be that roughly similar procedure will accommodate these different topics but what if the fix goes deeper than model behavior and extends to tech company business models or questions about the underlying data (as in the Ubuntu case)?

Decades of work in the sociology of science has shown that all attempts at democratic participation in science and technology [20], [21], [22] will inevitably favor certain types of participants and participation at the expense of others. Public hearings over nuclear power in the UK in the 70s, for example, by requiring certain kinds of evidence, sidelined the opinions of activist groups who wanted to make an economic rather than a safety argument. Focus groups and citizens assemblies favor professional participants (those who do it frequently) or those with the loudest voices. Often participation in science and technology favors facts, over say emotions or experiences. Despite our best efforts, there are no totally neutral participatory process, only ones which are more or less appropriate for particular issues.

Some of the teams realize this, for example, Common Ground (now Dembrane)'s Deliberation at Scale project report [23] remarks:

> We might need different tools for different situations… …It is our impression that democratic inputs to AI and AI contributions to democracy can take many different forms, should take many different forms and have to be developed in and with their local contexts. As a commentator on our initial proposal remarked "Socio-cultural specificity, not generalisability, might be a strength".

One of the teams, Inclusive AI, directly addressed this problem [24] testing different voting techniques (ranked voting, quadratic voting) on two different marginalized groups: people with disabilities in the US, and people from the Global South. The team used a modified Pol.is set up in which a Chatbot moderated focus group discussions around which values AI should follow. Then participants were asked to vote on which values to prioritize. While they did not find significant differences in 'user satisfaction' with different voting techniques between the two groups they did discover different levels of satisfaction with participants who were previously 'trusting' or 'distrusting' of AI technologies. This raises questions, however, about to what extent the use of AI moderation can be employed for those skeptical of AI. How are we to ensure that they are included in the debate? Perhaps these groups will need traditional human moderation. There is some evidence that OpenAI recognized this point after the workshops, noting that participants with anxieties about AI were perhaps less willing to participate.[5]

So there are clear problems with assuming one form of participation can suit everyone. Interestingly, Ron Eglash, a University of Michigan professor overseeing the Ubuntu project, who we spoke to over zoom, rejected the choice between local and specific and large-scale participation entirely. In a recent paper [25], the Ubuntu team proposes a 'fractal' understanding of participation in which distributed and organic interactions (branches) might feed into more centralized ones (trunk). In this model, different procedures accommodating different communities might still be synthesized together.

### 3d. Extracting Values

One of the most 'scalable' solutions, and the one employed by most of the teams is constitutional AI [26] which relies on a fine-tuned AI 'teaching' other AI models to ensure that the outputs are aligned to some desired behavior – defined through a written list of principles given by human experts. The AI is trained to follow these principles through a reinforcement learning process – in which the feedback is provided by a 'teacher AI' tasked with recognizing transgressions of principles. This process has had remarkable success in protecting AI outputs trained in the process against red-teaming attacks or attacks that can be classified more neatly into categories of well-documented societal harms, such as explicit manifestations of racist hate speech (among other examples).

Naturally then, most of the teams aimed to produce general statements or principles (or values) for AI to follow. This is probably because Open AI's call explicitly states that 'Laws encode values and norms to regulate behavior' suggesting that laws are what is needed [2]. This in line with AI Ethics and AI policy discussions which have amassed a host of values or principles like Fairness, Accountability, Transparency and Privacy which AI systems should aspire to.

It seems ideal to be able to simply give AI a list of agreed-upon principles to follow but it is not self-evident when a principle is being upheld or not or what to do when two different principles contradict [27]. As Carl Miller of the Recursive Public project put it to us: 'values or principles are a good starting point for discussions but on their own are not satisfying or efficient material for developing policy from.'

**Legal strategy for mobile game**

I run a popular online mobile game company. We have millions of users all across the United States. Our mobile games are free, but have microtransactions. Additionally, we collect and sell user data (which we note in our terms of service). Finally, we run ads in our games, which also generate revenue. What should our legal strategy include, consider, or be?

**Guest overstaying in apartment**

I let someone stay in my apartment as a favor, but now he won't leave. He doesn't have a job and is causing problems. His misbehaving daughter is here, too. When I asked him for money that he owes me, he hit me. My lease says guests can stay for 72 hours, but he's been here way longer. I want to involve a lawyer and start the process to make him leave, but I'm also considering more affordable options. What's the best thing to do?

**Bookstore board of directors**

I was recently invited to join a board of directors for my local bookstore. I've never served on a board before. What do I need to know about this service? Additionally, there are a few things that concern me. First, I enjoy patronizing a number of bookstores around the city. Second, my uncle owns one of these bookstores. Finally, I own a substantial amount of Barnes and Nobles stock. Should I have any concerns, legal or otherwise, with joining this board?

**User query: scope**

*"The user seems to be seeking advice on a process for eviction." — P9, postdoctoral researcher in law.*

**Legal: complexity**

*"There's potential criminal assault here too." — P9, postdoctoral researcher in law.*

**User attribute: geography**

*"What state/city is the user living in?" — P7, law student*

**Figure 3. Example of a case being modified along different dimensions [28].**

As one of the projects, <u>Case Law for AI [28], [29], [30]</u>, points out – it is incredibly hard for either AI (*or humans*) to know how to implement general statements. Values do not come with how-to instructions for each and every situation. 'Case law' refers to the idea that, rather than starting with universal principles (applicable across a country or region), principles can be built up from individual legal decisions [29]. The team developed a methodology in which participants are asked to select between model responses given a brief write-up of a situation. These specific 'cases' are generated by AI based on expert instructions for which 'dimensions' of the case might meaningfully alter the necessary actions (user demographics, place, severity). For example, 'a case' might be: how should an LLM respond was asked by a user what to do with a guest overstaying their welcome in an apartment? (Fig. 3) Given some initial text on the scenario, an AI can generate variations on the text if, say, the guest was male or female, or what US state the apartment is located in, or how long the transgression occurs for. By then showing these variations to participants, the team can build up a large repository of desired model responses given these slightly different scenarios.

The Case Law team sees their case-based solution as a supplement or qualifier for Constitutional AI principles, but it seems possible (and they confirmed this over email) that model retraining could be carried out entirely with cases. The only problem then becomes ensuring that enough cases can be generated in sensible ways and adjudicating between participants' preferences in particularly controversial or divisive situations.

The Case Law team was the exception however: most of the teams solicited abstract principles from their participants. It is also worth noting that these were generally normative statements about how LLMs should or should not behave. That is, the teams were asking participants for *solutions*, as opposed to asking what *problems* or *concerns* they have about AI. One team relayed to us that Open AI preferred statements in the form of principles (what an AI should or should not do), but there is nothing about these kinds of procedures which necessitates this. What is discussed between participants could be anything – a full policy document, problem statements, narratives of their experiences etc. So the choice to solicit values is likely conditioned by the requirements of Pol.is and of Constitutional AI, but there are many other types of inputs to consider.

### 3e. Consensus

Another common assumption shared across many of the teams (and another requirement of Constitutional AI) is that there is some method of agreeing on the principles to be adhered to. As already noted, Pol.is is based on voting and aims to promote consensus statements. Many of the teams, Ramesh [6], Inclusive AI [24] etc. are focused on developing better mathematical techniques for sorting statements and ensuring that people feel statements are fair.

One of the teams, the Meaning Alignment Institute [31], mentioned earlier, used a novel technique in which they asked people what they would do in specific scenarios, and then used an LLM Chatbot to press the participants on *why* they thought this was the best course of action. By repeatedly drilling down, the LLMs could move participants past knee jerk justifications and left-right slogans to the 'underlying values' which, they argue, transcend different political ideologies. This is an interesting solution to the problem of seeming disagreements over values, but as these values become more universal, we wonder if they might also become less useful for action. 'The wisdom of elders is important' may well be a principle shared by American Republicans and Democrats but it is unclear how this might be used to influence everyday LLM tasks.

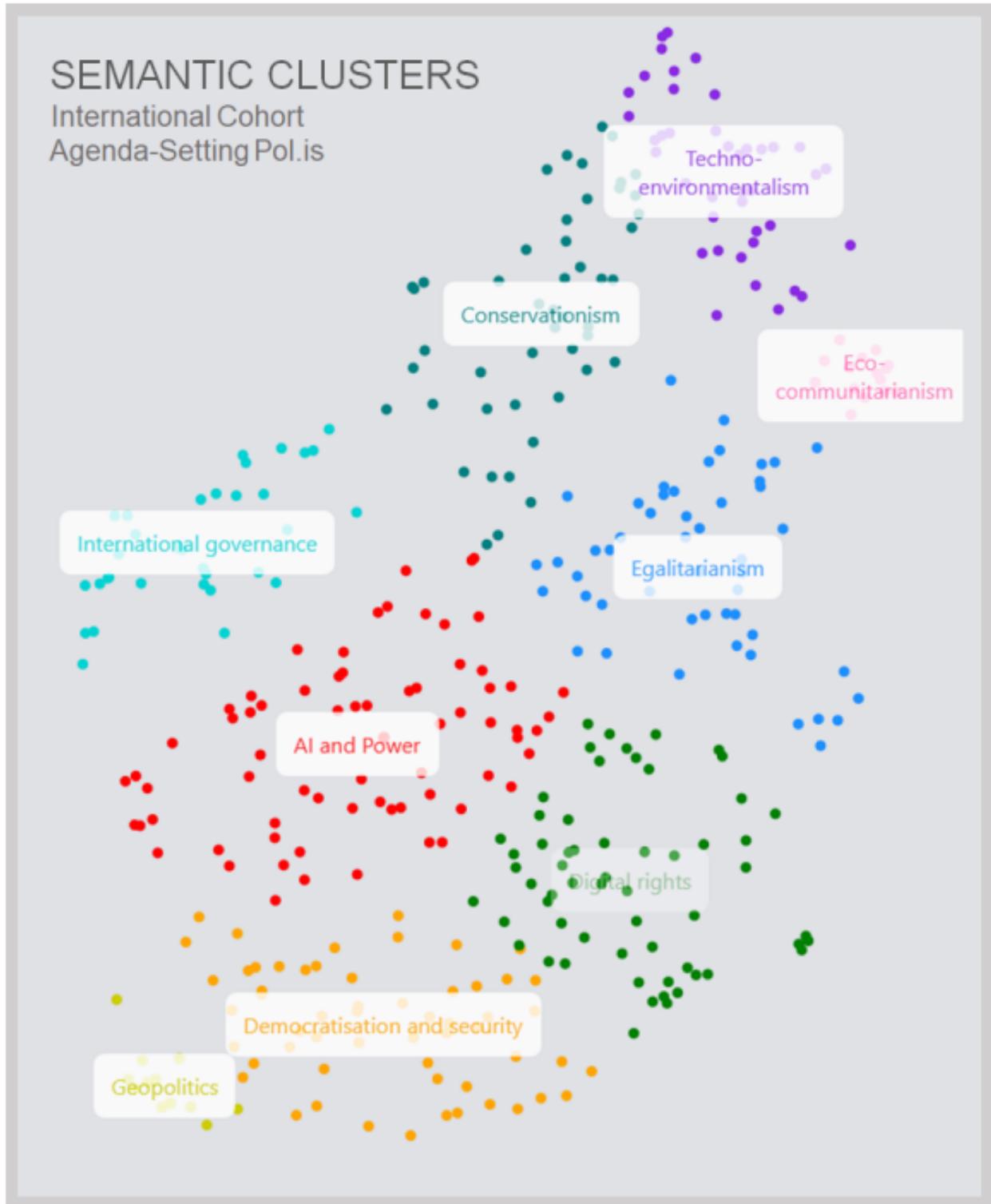

**Figure 4. "Semantic clustering of agenda-setting polis data."**

The Recursive Public project [32], led by Chatham House and vTaiwan, started from the premise that there might be a *danger* in over-emphasizing consensus and thus concealing the variety of possible opinions. Rather than using Pol.is to extract popular or consensus statements, the team used LLMs to thematically group related statements and display them as a colorful diagram (Fig. 4). This fractious landscape may not seem very helpful for implementing Constitutional AI, but as representatives of vTaiwan noted at a workshop earlier this year, they only use statements or principles as an elicitation device for in person negotiations, a starting point rather than an end point. As Carl Miller, a Chatham House-affiliated participation expert who worked on the project, pointed out 'The [participation] process should ideally be undulating in and out from the general to the particular." In other words, the movement from individual votes to a winner or from individual opinions to consensus need not be a one-way movement but a more recursive process. This is another way in which care and scale need not be 'either-or.'

Eglash (Ubuntu) sees consensus and diversity of opinions as a spectrum and argues that participation should aim to find a balance (between everyone's voice being heard and having actionable results). In the Fractal Governance article [25], they propose a metric – related to how many branches a conversation has – as a way of capturing when conversations sprawl outwards as opposed to returning to central points.

To their credit, Open AI seems to have acknowledged this tension in their response to the interim reports. As they put it 'It's not just about siding with the majority, but also giving a platform to different viewpoints.'[5]

### 3f. Representativeness

Open AI define democratic processes as '…a process in which a broadly representative group of people exchange opinions, engage in deliberative discussions, and ultimately decide on an outcome via a transparent decision making process' [2]. It is worth unpacking this definition.

Firstly, regarding 'transparecy' the teams have gone to great lengths to ensure that the deliberation process is seen as fair and 'transparent' for those participating (most offering a post-participation survey to gather participant's thoughts on the experience), but it is hard to say if a larger exercise could be 'transparent' with a corporation who does not disclose which data they used to train the models or release details of their, training, 'retraining', adaptation or 'guardrail ' procedures. OpenAI is a private company so we are not expecting them to open up their proprietary processes and datasets and there are also real dangers of jailbreaking (tricking LLMs into doing prohibited or illegal things) which could based on knowledge of the inner workings of a model. So there are good reasons not to disclose everthing. Still, it is worth asking if something essentially democratic is lost when participants cannot query what is under the hood.

Secondly, OpenAI makes clear that these experiments are not 'binding' for now: this was just a pilot study but there may be a process in future which is binding in which LLMs are regularly tweaked through feedback with publics through similar procedures [2]. Yet as Marci Harris, from Pop Vox pointed out (link) in a presentation at the TICTeC, Civic Tech conference in London, it is (by definition) not democratic without being binding. Without something at stake, these exercises are merely 'market research' as she provocatively put it. OpenAI at least acknowledges this in the call, citing a paper [33] which refers to 'participation washing' – or using narrow forms of participation to stave off meaningful regulation.

Thirdly, no democratic process regarding a technology can be considered complete without considering the underlying economic model and questions of ownership (of data as well as key infrastructure). Again, we bring this up not because we expect OpenAI to have made some binding agreement at this stage but because framing it as a mere 'pilot' has the effect of occluding more profound discussions of what the ultimate procedure might look like. It is notable that the Ubuntu team was the only one that tried to think through a procedure by which the specific affected community – artists in the African Continent – would have some accountability and ownership over model outputs.

Perhaps underlying these points, there seems to be an assumption that scalable solutions must involve *representative* publics, that a small group of participants must stand for the whole and that some form of voting will be necessary. This is likely because they believe that it would be too costly and messy to organize users as distributed organizations or involve pre-existing, concerned communities. Only the Inclusive AI project specifically targeted groups (disabled communities and participants from the global south) who might not receive enough attention in representative samples. Consulting these groups is not representative in the statistical sense but the presence of these groups corrects gaps in representation (in the political sense) in tech development more generally.

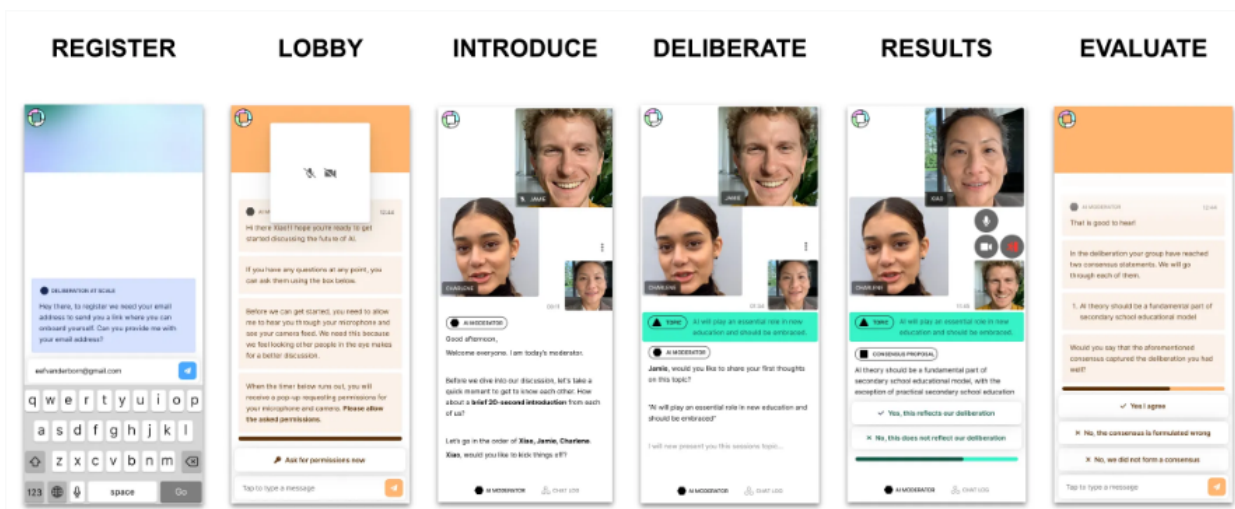

**Figure 5. A summary of Common Ground (now Dembrane)'s process [23].**

The other possibility discussed by some of the teams is that participation would not be in the service of self-regulation by tech companies but rather government regulation. This was important for both Recursive Publics and Common Ground. The Common Ground team's main innovation was to propose that an essential element of participatory exercises is their social dimension [23]. Different things happen when propositions are tested through small group discussions, rather than merely voted on in private. So they organized AI facilitated video calls between groups of 3-4 participants who decided on their principles collectively (Fig. 5). While our informants admitted the chatbot facilitators could be clunky at times, this didn't seem to matter for the participants who, on the whole, seemed to enjoy the process and claimed to have learned from it.

These participants were from a representative sample and randomly assigned to each other but there is no reason this procedure couldn't be applied to a geographically co-located group of users or joint owners as Common Ground confirmed. They cautioned however that there are pitfalls in completely ceding control over to publics. We cannot expect people to be well-informed both about AI and about how government works, so in line with the Recursive Publics project, these procedures might have to be an initial offering, a conversation starter with local government.

## 4. Alternative Forms of Participation

We've seen in the previous section that there are several, more-or-less taken for granted assumptions underlying the Democratic Inputs project which are worth questioning. So what would it look like to do participation differently: 1) to undertake small scale participation; 2) to develop models for specific use cases with specific communities; 3) to develop bespoke participation exercises; 4) to extract something other than principles; 5) to privilege discovery over consensus 6); and engage non-representative samples in a more accountable way?

We recently conducted a small, informal focus group with teachers in India, as a possible pilot for our own participatory exercise. Our starting gambit was that participation might unfold in very different ways if, like the Ubuntu team, we focused on a particular community and thus particular use cases and potential problems with Generative AI. We settled on AI in teaching and sourced some willing test subjects in Kolkata, West Bengal, India sourced through our personal networks. Naturally there have been many fears about LLMs helping students to cheat on assignments and write essays for them but there are also opportunities in education for LLMs to deliver more personalized teaching, delivering content at different reading levels.

As part of focusing on specific use cases, we felt it would be important for the teachers to actually use an LLM to undertake a practical task in the exercise (we used ChatGPT 4.0) rather than speculating about what it could do or relying on idealized trolley problems. After giving a brief explanation of LLMs and how they work, we asked the teachers to agree on a prompt to give to Chat GPT so that we could collectively discuss the response. The teachers considered a few possibilities before asking GPT to generate sample multiple choice quiz questions on a particular topic.

Chat GPT generated a few quizzes for math, physics and English. However, based on our previous experiments with Chat GPT, we suspected that it would fail to tailor prompts to local circumstances, where not enough data was available – relying on stereotypical or Orientalist caricatures of a culture or region. So we asked the teachers if questions like 'Identify the part of speech in parentheses in the following sentence "The (quick) brown fox jumps over the lazy dog."' were too culturally specific.

> Moderator: I'm curious, like, do you have foxes where you are? And y'know, is that like a problem for students if.. the examples are very Western or American or British

> Teacher: No I don't think that will be a problem because we try to expose them to lots of scenarios. If they need an explanation, I can do that also. No, that shouldn't be a barrier.

> Moderator: Would you like to see it try to make it more relevant to where you are.. ?

At our request, Chat GPT then replaced the question with 'Identify the part of speech of the word in parenthesis in the sentence "Durga Puja is (celebrated) with great enthusiasm in Kolkata."'

> Moderator: What do we think about this response?

> [long pause with some chuckling]

> Teacher: I think these kinds of sentences will definitely get them smiling [laughing].

> Moderator: Smiling because…?

> Teacher: Smiling because it's something happening around them.

Exchanges like this in participation exercises are difficult to interpret. Were the teachers laughing because they found it charming, or were they laughing at the bluntness of the attempt at tailoring? In either case, they seemed to say that GPT could be 'engaging' for students (though perhaps not for the right reasons). We have doubts if current LLMs tasked with summarizing focus group transcripts, could catch this sort of ambivalence, which relies on non-verbal cues and silence. In addition, as one of the Democratic Inputs teams noted, LLMs were not as effective as human moderators because they were too obsequious and deferential to ask tough follow up questions.

After seeing what Chat GPT could do, we then asked them what problems Chat GPT might cause in their industry. One teacher raised the concern that if GPT was producing similar quiz questions (or similar types of questions) across the country, then students could start to predict what would be on the quiz. Would it eventually run out of question types, or start to favor certain ones over others? How could teachers avoid this or coordinate such tasks together?

Our presumption that representational harms would be primary was probably the result of our being two cosmopolitan researchers, steeped in Fair/ML and Science and Technology Studies literature. Instead, it was more practical matters having to do with a specific task – quiz generation – which was the problem for the teachers, and it was only identified through the practical exercise of playing with GPT, followed by an open-ended discussion.

What this highlights is that when one does not limit in advance the problem definitions or move too quickly to developing consensus statements, the potential of deliberation exercises is for our participants to *surprise* us. Indeed, this should be the goal, if participation is not merely a means for convincing people or validating decisions already made. The most important possibility, when this problem-formulation remains open, is that participants might see AI itself *as a problem*, and realize they do not need it in their communities at all. There is always a danger that participation in AI, even while critiquing it, cements AI's inevitability and apparent usefulness [34]

Yet is this more open-ended type of participation limited to small scale interactions? Not necessarily. For example, Pol.is type voting systems, paired with group discussions could still work with more focused communities. But they could be based around practical tasks – getting the AI to do something. Then instead of soliciting values or principles we could solicit potential problems, fears or challenges. We could even solicit experiences people have had with similar technologies and get participants to vote on which ones resonate with them.

There is no shortage of models of alternative participation exercises to draw on. For example in relation to carbon capture and removal technologies, researchers have found innovative ways to describe technologies without making them sound inevitable (not using sleek promotional images for example) or to limit the use of (convincing sounding) expert jargon [35]. Similarly in relation to AI, Stilgoe and Cohen [34] try to conceive of a public dialogue which does not presume the end goal as 'acceptance' of the technology as proposed.

Forlano and Mathew [36] describe a host of approaches from speculative design which are applied to participation in urban informatics (smart cities), which include map-making, simulated design processes, brainstorming and prototyping. This resulted in the participants not only imagining alternative futures but amassing alternative values to the normal battery of 'efficiency', 'growth' presumed by technological development.

There are also many different possible outputs other than principles for Constitutional AI. For example, the outcome of participation experiments could be benchmark datasets consisting of desirable LLM responses (e.g. BBQ [37], PALMs [38], StereoSet [39] or, more generally, curated datasets of desirable behaviors as in the case of Inverse Reinforcement Learning (IRL). This would still require, though, that there are procedures for collectively deciding which behaviors or which datasets are desirable. The point is that the outputs could be any of these things – and that depends on the needs of the specific community and the problem at hand.

Finally, the participants need not be statistically representative – they could also be chosen representatives of specific communities or user groups and have a stake in the outcome. A number of initiatives such as data cooperatives, intermediaries and indigenous trusts have been conceptualized (for example TKLabels and Sahamati) which follow a community ownership model. Aligning the models through an RLHF (Reinforcement Learning with Human Feedback) process could then take place within these community actors/civil society entities and the alignment process could be specific to each community's needs and also specific to their model use case.

We could also involve participants in the design of new bespoke tools. Instead of retraining, which again, assumes a general model as standard, we could start with open-source alternatives like ROBERTA (Question Answering), Llama (Visual Question Answering), OpenVoice (text to speech) among many others and heavily tailor them (rather than tweak) to specific tasks, like quiz generation. The advantage of using open-source alternatives would be that participants could more transparently see the impacts of their design choices, and they could have real ownership over the entire pipeline, data, algorithms, outputs. Hopefully, with ownership will come more robust and enthusiastic participation.

## 5. CONCLUSIONS

In the months that have followed the release of the interim reports, it seems that Open AI has moved forward internally with their alignment research, inviting some of the teams to collaborate with them. Most of the teams we spoke to were quite happy with their experience on the project, they were able to interact with a lot of other 'participation nerds' – as one participant put it, and OpenAI was admirably hands off, once the projects were funded. By all accounts, they did not try to steer things.

However, many of the teams also had the impression that OpenAI was not all that interested in the ultimate results – some suspected that OpenAI already had a plan on how to proceed. Each of the teams are pursuing this work in different ways. Many have started to take these participatory experiments out into the world and apply them to topics which are not about AI at all (Common Ground/Dembrane and Meaning Alignment for example).

Open AI has hosted multiple remote talks with the participants on the Open AI Forum [8] which reiterate the findings of the interim reports – but we have yet to hear a clear articulation of what their approach to collective alignment will be. It does seem clear that they have involved members of the Case Law and Ramesh teams and possibly others. We reached out to OpenAI but have not yet received a response.

In this commentary piece, we identified six interrelated assumptions about participation present in the Democratic Inputs project 1) that participation must be scalable 2) that participation should be around a single model 3) that there should be one form of participation 4) that the goal is to extract values 5) that the values need consensus 6) that participants should be representative. Trying to solicit participation from representative publics is a good start, because it starts to correct the assumption of a universal user experience of LLMs. But the experience of smaller, marginal groups will *remain* marginal in these exercises, silenced by larger voting blocs or spoken to in the wrong language or approached with presumed concerns. If the model developers or owners decide the model purpose beforehand, or worse, buy into the myth of a general-purpose model, such a process of alignment will be participatory in name only.

Ultimately, we should not expect OpenAI, now a partially-profit making entity, to police or regulate itself or invite meaningful oversight from its users. As Jack Stilgoe remarked in a recent editorial [40], tech companies are only interested in design fixes; an entirely different process would be necessary to tackle AI in a wider sense (not just particular products). For Stilgoe this would require something like citizens assemblies, which would aim to discover people's real concerns in an open-ended way. For Eglash this might have to start more organically at the local level and grow fractally into something larger and more complex.

It may be the case that these exercises are, by design, tokenistic: a sort of smokescreen to prevent meaningful regulation from outside, but to the extent that these attempts at participation are earnest (and we do not doubt the good intentions of the teams and the facilitators at OpenAI) we need to think more carefully about what sort of participation we need. What is the point of participation? Who is it for?

In reviewing some of the apparent assumptions of the Democratic Inputs project our aim is to draw out the many ways in which participation in AI could be drastically different from the version which OpenAI proposed. We 1) need not trade care for scale; we could 2) focus on specific use cases instead of training a generic foundation model; 3) tailor participation to specific communities with particular needs; 4) discuss problems and experiences not just principles 5) explore the variety of opinions and 6) entertain models of joint ownership.

And importantly, as the teams have shown, none of these possibilities are precluded just because of the requirements for something scalable. Using LLMs for focus group moderation or to summarize statements still needs to be refined, but these experiments have shown that the basic set up of these exercises can work and can scale.

As we said, the technical possibilities employed by the Democratic Inputs teams are fascinating and innovations in the scaling of deliberation have an important role to play, but there needs to also be more innovation in the sorts of *questions we ask* and the way we engage people. We need to think of different ways, not just of scaling participation but of bringing participation *down* to scale.

## Acknowledgements

This work was supported by the project REIMAGINE ADM (https://chanse.org/reimagine-adm/) supported by the CHANSE ERA-NET Co-fund programme, which has received funding from the European Union's Horizon 2020 Research and Innovation Programme, under Grant Agreement no 101004509 and hosted by DiverseAI (https://www.diverse-ai.org/).

## Declaration of Interests

## References

[1]  J. Edelman and O. Klingefjord, 'OpenAI x DFT: The First Moral Graph', Meaning Alignment Institute. Accessed: Mar. 05, 2025. [Online]. Available: https://meaningalignment.substack.com/p/the-first-moral-graph

[2]  Open AI, 'Democratic inputs to AI', OpenAI.com. Accessed: Nov. 04, 2024. [Online]. Available: https://openai.com/index/democratic-inputs-to-ai/

[3]  B. Perrigo, 'Inside OpenAI's Plan to Make AI More "Democratic"', TIME. Accessed: Nov. 04, 2024. [Online]. Available: https://time.com/6684266/openai-democracy-artificial-intelligence/

[4]  C. Small, M. Bjorkegren, T. Erkkilä, L. Shaw, and C. Megill, 'Polis: Scaling Deliberation by Mapping High Dimensional Opinion Spaces', *Recer. Rev. Pensam. Anàlisi*, vol. 26, no. 2, 2021.

[5]  OpenAI, 'Democratic inputs to AI grant program: lessons learned and implementation plans', OpenAI.com. Accessed: Nov. 04, 2024. [Online]. Available: https://openai.com/index/democratic-inputs-to-ai-grant-program-update/

[6]  A. Konya, L. Schirch, C. Irwin, and A. Ovadya, 'Democratic Policy Development using Collective Dialogues and AI', Nov. 03, 2023, *arXiv*: arXiv:2311.02242. Accessed: Oct. 11, 2024. [Online]. Available: http://arxiv.org/abs/2311.02242

[7]  P. Olson, 'Big Tech Has Our Attention — Just Not Our Trust', *Bloomberg.com*, 2024. Accessed: Jan. 27, 2025. [Online]. Available: https://www.bloomberg.com/graphics/2024-opinion-ai-solution-big-tech-trust-problem/

[8]  OpenAI, 'Virtual Event: Collective Alignment: Enabling Democratic Inputs to AI - Event', OpenAI Forum. Accessed: Jan. 08, 2025. [Online]. Available: https://forum.openai.com/home/events/collective-alignment-enabling-democratic-inputs-to-ai-2024-04-18?agenda_day=6601afa6233573d487bf510f&agenda_track=6601afa7233573d487bf5121&agenda_stage=6601afa6233573d487bf5115&agenda_filter_view=stage&agenda_view=list

[9]  B. Wynne, 'Risk as globalizing 'democratic'discourse? Framing subjects and citizens', *Sci. Citiz.*, pp. 66–82, 2005.

[10] B. Glaser and A. Strauss, *The discovery of grounded theory: Strategies for qualitative research*. Aldine de Gruyter, 1967.

[11] J. Deville, M. Guggenheim, and H. Zuzana, *Practising Comparison: Logics, Relations, Collaborations*. Mattering Press, 2016.

[12] I. Tavory and S. Timmermans, *Abductive analysis: Theorizing qualitative research*. University of Chicago Press, 2014.

[13] N. Seaver, 'Care and Scale: Decorrelative Ethics in Algorithmic Recommendation', *Cult. Anthropol.*, vol. 36, no. 3, Art. no. 3, Aug. 2021, doi: 10.14506/ca36.3.11.

[14] A. Mol, I. Moser, and J. Pols, Eds., *Care in Practice: On Tinkering in Clinics, Homes and Farms*. transcript Verlag, 2010. doi: 10.1515/transcript.9783839414477.

[15] A. Hanna and T. M. Park, 'Against Scale: Provocations and Resistances to Scale Thinking', arXiv.org. Accessed: Oct. 10, 2024. [Online]. Available: https://arxiv.org/abs/2010.08850v2

[16] K. Zhou *et al.*, 'Don't Make Your LLM an Evaluation Benchmark Cheater', Nov. 03, 2023, *arXiv*: arXiv:2311.01964. Accessed: Jul. 15, 2024. [Online]. Available: http://arxiv.org/abs/2311.01964

[17] A. Caliskan, J. J. Bryson, and A. Narayanan, 'Semantics derived automatically from language corpora contain human-like biases', *Science*, vol. 356, no. 6334, pp. 183–186, Apr. 2017, doi: 10.1126/science.aal4230.

[18] H. Suresh, E. Tseng, M. Young, M. Gray, E. Pierson, and K. Levy, 'Participation in the age of foundation models', in *The 2024 ACM Conference on Fairness, Accountability, and Transparency*, Rio de Janeiro Brazil: ACM, Jun. 2024, pp. 1609–1621. doi: 10.1145/3630106.3658992.

[19] M. Nayebare, R. Eglash, U. Kimanuka, R. Baguma, J. Mounsey, and C. Maina, 'Interim Report for Ubuntu-AI: A Bottom-up Approach to More Democratic and Equitable Training and Outcomes for Machine Learning', Sep. 2023.

[20] B. Wynne, *Rationality and ritual*. Abingdon: Earthscan, 2011.

[21] A. Irwin and M. Michael, *Science, Social Theory and Public Knowledge*. Maidenhead, UK: Open University Press, 2003.

[22] J. Stilgoe, S. J. Lock, and J. Wilsdon, 'Why should we promote public engagement with science?', *Public Underst. Sci.*, vol. 23, no. 1, pp. 4–15, 2014.

[23] Dembrane, 'First Report: Democratic Inputs to AI', Dembrane.com. [Online]. Available: https://www.dembrane.com/blog/report-openai-october-2023

[24] T. Sharma *et al.*, 'INCLUSIVE. AI: ENGAGING UNDERSERVED POPULATIONS IN DEMOCRATIC DECISION-MAKING ON AI', Accessed: Oct. 10, 2024. [Online]. Available: https://socialcomputing.web.illinois.edu/images/Report-InclusiveAI.pdf

[25] R. Eglash *et al.*, 'AI governance through fractal scaling: integrating universal human rights with emergent self-governance for democratized technosocial systems', *AI Soc.*, Jul. 2024, doi: 10.1007/s00146-024-02029-4.

[26] Y. Bai *et al.*, 'Constitutional AI: Harmlessness from AI Feedback', Dec. 15, 2022, *arXiv*: arXiv:2212.08073. Accessed: Jan. 12, 2024. [Online]. Available: http://arxiv.org/abs/2212.08073

[27] J. Dewey, *The quest for certainty*. Minton, Balch, 1929.

[28] Q. Z. (Jim) Chen, K. Feng, I. Cheong, A. X. Zhang, and K. Xia, 'Case Law for AI Policy - Project Website'. Accessed: Nov. 04, 2024. [Online]. Available: https://social.cs.washington.edu/case-law-ai-policy/

[29] K. J. K. Feng, Q. Z. Chen, I. Cheong, K. Xia, and A. X. Zhang, 'Case Repositories: Towards Case-Based Reasoning for AI Alignment', Nov. 26, 2023, *arXiv*: arXiv:2311.10934. Accessed: Oct. 11, 2024. [Online]. Available: http://arxiv.org/abs/2311.10934

[30] Q. Z. Chen and A. X. Zhang, 'Case Law Grounding: Using Precedents to Align Decision-Making for Humans and AI', Sep. 06, 2024, *arXiv*: arXiv:2310.07019. doi: 10.48550/arXiv.2310.07019.

[31] O. Klingefjord, R. Lowe, and J. Edelman, 'What are human values, and how do we align AI to them?', Apr. 17, 2024, *arXiv*: arXiv:2404.10636. Accessed: Oct. 11, 2024. [Online]. Available: http://arxiv.org/abs/2404.10636

[32] F. Devine *et al.*, 'Recursive Public: Piloting Connected Democratic Engagement with AI Governance', 2023.

[33] M. Sloane, E. Moss, O. Awomolo, and L. Forlano, 'Participation is not a Design Fix for Machine Learning', Aug. 11, 2020, *arXiv*: arXiv:2007.02423. doi: 10.48550/arXiv.2007.02423.

[34] J. Stilgoe and T. Cohen, 'Rejecting acceptance: learning from public dialogue on self-driving vehicles', *Sci. Public Policy*, vol. 48, no. 6, pp. 849–859, 2021.

[35] R. Bellamy and J. Lezaun, 'Crafting a public for geoengineering', *Public Underst. Sci.*, vol. 26, no. 4, pp. 402–417, May 2017, doi: 10.1177/0963662515600965.

[36] L. Forlano and A. Mathew, 'From design fiction to design friction: Speculative and participatory design of values-embedded urban technology', in *Urban Informatics*, Routledge, 2017, pp. 7–24. Accessed: May 21, 2024. [Online]. Available:

https://www.taylorfrancis.com/chapters/edit/10.4324/9781315652283-2/design-fiction-design-friction-speculative-participatory-design-values-embedded-urban-technology-laura-forlano-anijo-mathew

[37] A. Parrish *et al.*, 'BBQ: A Hand-Built Bias Benchmark for Question Answering', Mar. 15, 2022, *arXiv*: arXiv:2110.08193. Accessed: Jan. 22, 2024. [Online]. Available: http://arxiv.org/abs/2110.08193

[38] I. Solaiman and C. Dennison, 'Process for Adapting Language Models to Society (PALMS) with Values-Targeted Datasets', Nov. 23, 2021, *arXiv*: arXiv:2106.10328. doi: 10.48550/arXiv.2106.10328.

[39] M. Nadeem, A. Bethke, and S. Reddy, 'StereoSet: Measuring stereotypical bias in pretrained language models', Apr. 20, 2020, *arXiv*: arXiv:2004.09456. Accessed: May 21, 2024. [Online]. Available: http://arxiv.org/abs/2004.09456

[40] J. Stilgoe, 'AI has a democracy problem. Citizens' assemblies can help.', *Science*, vol. 385, no. 6711, p. eadr6713, Aug. 2024, doi: 10.1126/science.adr6713.